\documentstyle[12pt,epsf]{article}
\def\beq{\begin{eqnarray}}   \def\eeq{\end{eqnarray}}

\def\a{{\rm AdS_5}}

\begin{document}
\thispagestyle{empty}
\begin{flushright}
NYU-TH/00/02/02 \\
\end{flushright}

\vspace{0.1in}
\begin{center}
\bigskip\bigskip
{\Large \bf Metastable Gravitons and  Infinite Volume Extra Dimensions}

\vspace{0.3in}

{ G. Dvali, G. Gabadadze and  M. Porrati}
\vspace{0.1in}

{\baselineskip=14pt \it 
Department of Physics, New York University, New York, NY 10003 } \\
\vspace{0.2in}
\end{center}

\vspace{0.9cm}
\begin{center}
{\bf Abstract}
\end{center} 
\vspace{0.1in}

We address the issue of whether extra dimensions could  have an infinite
volume  and yet  reproduce the effects of observable four-dimensional 
gravity on a brane.
There is no normalizable zero-mode graviton in this case, nevertheless 
correct Newton's law can be  obtained by exchanging bulk
gravitons. This can be interpreted as an exchange of 
a single {\it metastable} 4D  graviton.
Such theories have remarkable  phenomenological signatures 
since the evolution of the Universe becomes  
high-dimensional at very large scales. 
Furthermore, the bulk supersymmetry in the infinite volume limit might  
be preserved  while being completely broken on a brane. 
This gives rise to a  possibility of controlling 
the value of the bulk cosmological constant.
Unfortunately, these  theories have difficulties  in  reproducing certain
predictions of Einstein's theory related to relativistic sources.
This is due to the van Dam-Veltman-Zakharov discontinuity in the propagator 
of a massive graviton. This suggests that all theories in which 
contributions  to 
effective 4D gravity come predominantly from the bulk graviton exchange
should encounter  serious phenomenological difficulties.
 
\newpage

If Standard Model particles are localized on a brane, 
the volume of extra dimensions
can be as large as a millimeter without conflicting to any experimental
observations \cite{add}. This is also true for warped spaces  in which
the extra dimensions are non-compact but have a finite volume 
\cite{RS} (for earlier works on warped compactifications see 
\cite {Visser,Gogber}).

In the framework of Ref.  \cite{add} the volume 
$V\sim L^N$ of extra $N$ space dimensions sets the normalization of
a four-dimensional graviton zero-mode. 
Therefore,  the relation between the
observable and the fundamental Planck scales ($M_P$ and 
$M_{Pf}$ respectively) reads as follows:
\beq
M_P^2~ = ~M_{Pf}^{2+N}V~.
\label{pl}
\eeq
A similar  relation holds for the Randall-Sundrum (RS) scenario \cite{RS}, 
where the role of $L^{-2}$ is played
by the curvature of $\a$. In this case the 
extra dimension is not compact, nevertheless its
length (or volume) is finite and is determined by the 
bulk cosmological constant $L\propto 1/\sqrt{|\Lambda|}$.    
In the scenario of   
Ref. \cite{add}  gravity becomes  high-dimensional  at 
distances $r << L$ with the corresponding change in Newton's law
\begin{equation}
 1/r \rightarrow 1/r^{1 +N}~.
\end{equation}
The same holds true for RS-type scenarios with
non-compact extra dimensions \cite{addk}.

The purpose of the present letter is to study whether 
the volume of extra space can be  truly
{\it infinite} while the four-dimensional Planck mass
is still finite. In this case the relation (\ref {pl}) 
should somehow be evaded.  
Since there are  no normalizable zero modes in such cases, 
the effects of 4D gravity must be reproduced
by exchanging the  continuum of bulk modes.
An example of this type was recently proposed in Ref. \cite {Rubakov}. 
The physical reason why
such an exchange can indeed mimic the 
$1/r$ Newtonian law can be understood as follows.
The four-dimensional graviton, 
although is not an eigenstate of the linearized theory,
can still exist as a {\it metastable} resonance
with a finite lifetime $\tau_g$.
In such a case one might  hope that the exchange 
of this  graviton approximates  Newton's law at
distances shorter than the graviton
lifetime  but changes  the laws of gravity 
at larger scales.
The question is whether the four-dimensional effective 
theory obtained in this way is phenomenologically viable.
In the present paper we will  argue that, despite the correct Newtonian limit, 
the infinite volume scenario
has problems in reproducing other predictions of Einstein's theory.
This problem is shared by any model in which the dominant contribution
to 4D gravity comes from the exchange of continuum states.
The reason behind such a discrepancy is the  van Dam-Veltman-Zakharov discontinuity
in the propagator of a massive spin-2 field in the massless limit \cite  
{Veltman}. Very briefly, the physical effects of additional polarizations
of a massive graviton survive in the massless limit and change dramatically 
predictions of the theory \cite {Veltman}.  
As a result, any theory which relies on the exchange of massive
(no matter how light) gravitons, will give rise to predictions which 
differ  from those of General Relativity.

We shall consider the five-dimensional Einstein gravity coupled
to an arbitrary energy-momentum tensor $T_{AB}$  which is independent of four
space-time coordinates $x_{\mu}$. We will assume that $T_{AB}$  results
from a certain  combination of branes, bulk cosmological constant $\Lambda$ 
and classical field configurations which preserve  four-dimensional Poincar\'e
invariance. Einstein's equations $G_{AB} = T_{AB}+g_{AB}\Lambda $
give rise to a metric of the following form
\begin{equation}
ds^2~ =~ A(z)~\left ( ds^2_4 - dz^2 \right)~,
\end{equation}
where $z$ is an extra coordinate, and we assume that the four-dimensional
metric $ds^2_4$ is flat.
The volume of the extra space in this construction is determined by
an integral
\begin{equation}
\int_{-\infty}^{+\infty} A^{5/2}(z)~dz~.
\label{volume}
\end{equation}
For instance, in the RS framework $A(z)~=(1+H|z|)^{-2}$,
where $H$ is proportional to the square
root of the cosmological constant $H\propto \sqrt{|\Lambda|}$.
This warp factor gives rise to a finite expression in (\ref {volume}).
An infinite volume theory is  
obtained if, for instance, the value of $A(z)$ tends to a  {\it nonzero}
constant as $z$ goes to $\pm \infty$.
If  the volume is finite, there is a normalizable  zero-mode graviton in the
spectrum of fluctuations about this background \cite {RS}. 
Indeed, let us parametrize
the four-dimensional graviton fluctuations as follows:
\begin{equation}
ds^2 = A(z)\left [(\eta_{\mu\nu} +  h_{\mu\nu}(x,z))dx^{\mu}dx^{\nu}  - 
dz^2\right ]~.
\end{equation}
The corresponding linearized Schr\"odinger equation for the excitation
$h_{\mu\nu}(x,z) = A^{-3/4}(z) \Psi (z) h^{(0)}_{\mu\nu}(x_{\mu})$ 
takes the form:
\beq
\left ( -\partial_z^2  +  {3\over 4}\left \{  {A^{\prime\prime} \over A}
-{A^{\prime 2} \over 4 A^2}    \right \} \right )~\Psi (z)~=~m^2~\Psi(z) ~.
\label{h}
\eeq
Where $\eta^{\mu\nu} \partial_\mu\partial_\nu h^{(0)}=-m^2h^{(0)}$
and primes denote differentiation with respect to $z$.
This equation has a zero-mode solution
for a generic form of a  warp factor $A(z)$:
\begin{equation}
\Psi_{\rm zm}(y) ~=~A^{3/4} (z)~.
\end{equation}
This implies that the $z$ dependent part of the fluctuations 
$h(x,z)$ is just a constant, 
i.e., $h(x,z)= {\rm const.}{\rm exp}(ipx)$.
If the integral in (\ref{volume}) diverges, the volume is infinite and
the zero-mode is not normalizable.  
Thus, the spectrum in this case would consist of continuum states only.
One should notice, however, that  
even in this  case the correct Newtonian limit may be recovered 
in some approximation by exchanging
the continuum of non-localized states!  
The physical reason for such a behavior
is that the 4D localized graviton, although not an eigenstate, 
can still exist
as a {\it metastable} resonance with an exponentially long lifetime, $\tau_g$.
Therefore, the exchange of this resonance could  give rise
to a correct Newtonian potential at intermediate distances.
Note that this is just a reformulation of the fact that 
the tower of the bulk states conspire in such a way that the 
Newtonian potential is seen as an approximation.  
To make these discussions a bit precise let us turn to the 
propagator of a massive metastable graviton. 
This is given as follows:
\begin{equation}
G_4^{\mu\nu\alpha\beta}(x)=\int {dp^4\over (2\pi)^4} \left ( {1\over
2}(g^{\mu\alpha}g^{\nu\beta} + g^{\mu\beta}g^{\nu\alpha}) - {1\over
3}g^{\mu\nu}g^{\alpha\beta}+{\cal O}(p) \right )D(p^2,m_0^2, \Gamma)
e^{-ip(x)},
\end{equation}
where $D(p^2,~m_0^2, ~\Gamma)$ stands for the  scalar part of a
massive resonance propagator, $D(p^2,m_0^2, \Gamma)=
(p^2-m_0^2+im_0\Gamma)^{-1}$. $\Gamma$ denotes the width of the resonance.
The momentum dependent part in the tensor structure 
gives zero contribution when the propagator is convoluted with a 
conserved energy-momentum  tensor, thus, this part will be omitted. 
In order to make a contact with 
continuum modes, let us use the following spectral representation
for $D$:
\beq
D(p^2,~m_0^2, ~\Gamma)~=~{1\over 2\pi} 
\int~ { \rho(s)  \over s - p^2+i\epsilon  }~ds~,
\label{disp}
\eeq 
where $s$ denotes the Mandelstam variable and 
$\rho (s) $ is a spectral density. If the assumption of the 
resonance dominance is made, then $\rho(s)$ is approximated by  
a sharply peaked function
around the resonance mass  $s=m_0^2$. 
In what follows we assume that the resonance lifetime is 
very big and we neglect the effects of a  nonzero resonance width
(these will modify gravity laws at very large distances only).  
Exchanging  such a particle between  two static sources one obtains
the potential
\begin{equation}
V(r) \sim \int ~{e^{-\sqrt{s}r} \over r}~\rho(s)~ds~.
\end{equation}
This expressions reproduces a 
standard $1/r$ interaction at distances $r << (m_0)^{-1}$
in the single, narrow-resonance approximation, $\rho(s) \propto  
\delta(s - m_0^2)$.
On the other hand, we can expand the spectral density 
$\rho(s)$ into the complete set of bulk modes
\begin{equation}
\rho(s) = \int_0^{\infty}~ |\psi_m(0)|^2 ~\delta~(s - m^2)~dm~.
\end{equation}
Here, $\psi_m(0)$ denote the wave functions 
of the bulk modes at the point $z=0$. 
Using this expression for the spectral density one finds the 
following potential 
\begin{equation}
V(r) ~\sim ~\int_0^{\infty} ~{e^{-mr} \over r} ~|\psi_m(0)|^2~dm~. 
\label{continuum}
\end{equation}
This is nothing but the potential mediated by the continuum of 
the bulk modes. Thus, the effect of
the metastable graviton, when it exists, 
can be read off the expression which includes all the 
bulk modes. These two descriptions are complementary to each other. 
In the case when the resonance exists, the continuum modes 
can conspire in such a way that (\ref {continuum}) yields the
$1/r$ law in a certain approximation. The inverse statement 
is also likely to be true.   
In the appendix we will show explicitly  
the presence of a resonance state in a model with infinite 
volume extra dimension and $1/r$ potential produced
by the bulk modes \cite {Rubakov}.  As we mentioned above, 
such model gives Newtonian gravity 
only at intermediate distances. At large distances, the
five-dimensional laws of gravity should be restored
(due to the metastable nature of the resonance). 
This phenomenon could have dramatic cosmological and 
astrophysical consequences. Indeed, at large cosmic scales 
the time-dependence of the scale factor $R(t)$ in Freedman-Robertson-Walker
metric would dramatically change due to the change in the 
laws of gravity\footnote{A different possibility to modify the long
distance gravity due to an additional massive graviton  
was proposed earlier in
Ref. \cite {Kogan}. In the view of phenomenological 
problems discussed below, 
this graviton should be very weakly coupled.}.

Another interesting comment concerns bulk supersymmetry. 
Since the volume of the extra dimension
is infinite, it might be  possible to realize the following scenario.
The bulk is exactly supersymmetric and SUSY  is completely broken on 
a brane (this could be a non-BPS brane which is stable for some 
topological reasons \cite {DS}). The transmission of  SUSY breaking 
from the brane worldvolume to the bulk is suppressed by the 
volume of the bulk and is vanishing. In such a case one could imagine
a setup where the bulk cosmological constant is zero 
due to the bulk SUSY\footnote{Note that local SUSY does not necessarily
imply vanishing of the vacuum energy. However, this can be 
accomplished by imposing on a model additional global 
symmetries.}.
     
Having these attractive features of the theories with truly 
infinite extra dimensions discussed, we move to some phenomenological
difficulties of these models. In fact, 
we will argue below that these theories cannot reproduce other
predictions of Einstein's  general relativity.
The reason is that all the
spin-2 modes that dominantly contribute to the four-dimensional gravity 
in this case are massive modes.
It has been known for a long time \cite {Veltman} that propagator of 
massive spin-2  states has no continuous massless limit.
As a result the effects of the massless spin-2 graviton are different 
from the massive one, no matter how small the mass is.
Let us show how this affects the phenomenology of infinite volume theories.
The four-dimensional gravity on a  brane  is reproduced by an exchange
of the  continuum of bulk gravitons. At a tree level this gives
\begin{equation}
G_5\int_0^{\infty} dm \int dx^4~ d^4x'~
T_{\mu\nu}(x)~ G_m^{\mu\nu\alpha\beta}(x - x')~T'_{\alpha\beta}(
x') ~,
\label{exch}
\end{equation}
where $ T_{\mu\nu}(x)$  and $T'_{\mu\nu}(x')$ are the energy-momentum
tensors for two gravitating sources.
For $m\neq 0$  the graviton propagator is given by
\begin{equation}
G_m^{\mu\nu\alpha\beta}(x - x') = \int {dp^4\over (2\pi)^4} {{1\over
2}(g^{\mu\alpha}g^{\nu\beta} + g^{\mu\beta}g^{\nu\alpha}) - {1\over
3}g^{\mu\nu}g^{\alpha\beta}+{\cal O}(p)  \over p^2 - m^2 - i\epsilon} e^{-ip(x - x')}~,
\end{equation}
whereas for $m = 0$  we have
\begin{equation}
G_0^{\mu\nu\alpha\beta}(x - x') = \int {dp^4\over (2\pi)^4} {{1\over
2}(g^{\mu\alpha}g^{\nu\beta} + g^{\mu\beta}g^{\nu\alpha}) - {1\over
2}g^{\mu\nu}g^{\alpha\beta}+{\cal O}(p)  \over p^2 - 
i\epsilon} e^{-ip(x - x')}~.
\end{equation}
As we see, the tensor structures in the two cases are different. 
In the massless limit, the propagator exhibits the celebrated 
van Dam-Veltman-Zakharov
discontinuity. This is due to the difference in the number of
degrees of freedom
for massive and massless spin-2 fields.  
In our case this difference is very transparent, KK gravitons at
each mass level ``eat up'' three extra degrees of freedom of 
$g_{5\mu}$
and $g_{55}$ components of the higher dimensional metric
(``graviphotons'' and ``graviscalars'' respectively).

Since we choose a model in which there is no normalizable-zero 
mode, the whole answer is given by the bulk continuum.
Let us show that the 4D gravity which is 
obtained in this way cannot reproduce observable effects of 
General Relativity.
Let us first consider the Newtonian limit.  In this case, we take two static
point-like sources
\begin{equation}
T_{\mu\nu}(x) = m_1 \delta_{\mu 0}\delta_{\nu 0}
\delta(\vec {x}), ~~~T'_{\mu\nu}(x') =  
m_2 \delta_{\mu 0}\delta_{\nu 0}\delta(
\vec {x'} -
\vec {r})~.
\end{equation}
For this setup the bulk graviton exchange gives
\begin{equation}
{2\over 3}~m_1m_2 ~G_5~\int ~dm ~{e^{-mr} \over r} ~|\psi_m(0)|^2~.
\end{equation}
Since  the leading  behavior of the integral 
for the particular case at hand is $1/r$,
\begin{equation}
\int~ dm ~{e^{-mr} \over r}~ |\psi_m(0)|^2  \sim  {a\over r}  +  ...,
\end{equation}
the correct Newtonian limit may be reproduced 
($a$ is some normalization constant).  
On the other hand, since the exchange of one
normalizable  massless graviton would give
\begin{equation}
G_N {1\over 2} {m_1m_2 \over r}~,
\end{equation}
we have to set
\begin{equation}
a~G_5~=~{3~ G_N\over 4}~. 
\label{gnewton}
\end{equation}
This identification provides  the  correct Newtonian potential
for static sources. So far so good. 
Unfortunately,  the problem arises when one tries  to account for moving
sources.  To see this let us take
one of the sources to be a moving point-like  particle of mass $m_2$ and proper time 
$\tau$. The energy-momentum tensor for this particle is written as:
\begin{equation}
T'_{\mu\nu}(x') ~= ~m_2~\int ~d\tau ~\dot{x_{\mu}}~\dot{x_{\nu}} ~
\delta( x'- x(\tau))~.
\end{equation}
The result  of the bulk graviton exchange then gives
\begin{equation}
G_5~m_1m_2 ~\int ~d\tau ~(\dot{x_0}\dot{x_0} - {1\over 3}
\dot{x_{\mu}}\dot{x^{\mu}} )~\int ~dm ~{e^{-mr(\tau)} \over r(\tau)}
~|\psi_m(0)|^2~,
\end{equation}
where $r = \vec {x}(\tau)$.  
With the identification (\ref{gnewton}),
in the leading order this yields 
\begin{equation}
{3\over 4}G_Nm_1m_2 \int d\tau (\dot{x_0}\dot{x_0} - {1\over 3}
\dot{x_{\mu}}\dot{x^{\mu}} ) {1 \over r(\tau)}~. 
\end{equation}
On the other hand the exchange of a normalizable graviton zero-mode 
produces the following result 
\begin{equation}
G_Nm_1m_2 \int d\tau (\dot{x_0}\dot{x_0} - {1\over 2}
\dot{x_{\mu}}\dot{x^{\mu}} ) {1 \over r(\tau)}~. 
\end{equation}
This shows the  discrepancy between the predictions of the two theories. In
particular, the same procedure  applied
to the problem of bending of light by the Sun gives the discrepancy by the  
factor $3/4$. Indeed, for the bending of light in the gravitational field
of the Sun, the  tree-level bulk graviton exchange gives:
\beq
G_5~M_{\rm Sun}~T_{00}(k,q,\epsilon_\mu, \epsilon'_\nu)~\int~
dm~{\delta(k_0-q_0) \over (k-q)^2-m^2-i\varepsilon }~|\psi_m(0)|^2~\simeq
\nonumber \\
-{3\over 4} {G_N~ M_{\rm Sun}~T_{00} \delta(k_0-q_0)  
\over |\vec{k} -\vec{q}|^2}~+...~,
\eeq  
where $T_{00}(k,q,\epsilon_\mu, \epsilon'_\nu)$ is the component of 
the energy-momentum tensor for photons in the momentum representation,
and $k~(\epsilon_\mu)$ and $q~(\epsilon'_\nu)$ are the momenta(polarizations)
of initial and final photons. This is just $3/4$ of the result of the 
4D massless graviton exchange. 

Summarizing, we have shown that in theories with 
truly infinite extra dimensions
the correct four-dimensional Newtonian gravity
can be obtained at intermediate distances due to a metastable resonance 
graviton. This description is complementary to the exact summation of
continuum modes. Due to the finite lifetime of the resonance 
the laws of gravity are modified at large distances. This would  give rise
to interesting cosmological consequences. 
Moreover, these models could  allow to preserve bulk supersymmetry 
while it is completely broken on a brane.
Unfortunately, these models encounter a number of
phenomenological difficulties. The effects of additional polarization   
degrees of freedom of massive gravitons survive even in the 
massless limit and lead to  substantial discrepancies
with the predictions of General Relativity. 
It might be possible to cure these discrepancies   
by introducing new very unconventional interactions.
The addition of dilaton-type scalars coupled to
$T_{\mu}^{\mu}$ seems to make things worse.

\vspace{0.5cm}

{\bf Acknowledgments}

\vspace{0.2cm} 

We wish to thank Ian  Kogan and Valery  Rubakov for comments. 
The work of G.D. is supported in part by David and Lucile  
Packard Foundation Fellowship for Science and Engineering 
and  by Alfred P.  Sloan Research Fellowship. That of G.G. is
supported by  NSF grant PHY-94-23002. 
M.P. is supported in part by NSF grant PHY-9722083.

\vspace{0.5cm}

{\bf Note added}

\vspace{0.1cm} 

After this paper was prepared for submission  
the work \cite {Csaki} appeared.
The authors of this work have also realized that 
metastable gravitons can be responsible for the $1/r$ 
law in theories with infinite extra dimensions. However,
the generic phenomenological difficulties of this  
class of theories which is a  crucial part of our work have  not 
been addressed  in \cite {Csaki}. 

After this work appeared on the net, we were informed by V. Rubakov that
in the revised version of Ref. \cite {Rubakov} 
the role of a metastable graviton 
was also elucidated and its decay width was calculated.

\vspace{0.2in}
\begin{center}
{\bf Appendix}
\end{center}
\vspace{0.2in}

Below  we show the presence of a resonance
in a  system with an infinite extra dimension.
The particular example which we consider  
is the one studied in  Ref. \cite {Rubakov}.
The five-dimensional interval defining the background and 
four-dimensional graviton fluctuations  is set as follows (here we 
choose to use a  non-conformaly flat metric in order 
to be consistent with the conventions of \cite {Rubakov}):
\beq
ds^2~=~\left [ A(y) \eta_{\mu\nu}~+~h_{\mu\nu}(x,y) \right ] dx^\mu dx^\nu~-~
dy^2~.
\eeq
There is a  3-brane with positive tension $T$ which  is located at $y=0$.
In addition, there are two 3-branes with equal negative tensions 
$-T/2$ located at a distance $y_c$ to the left and right of 
the positive-tension brane. 
For $|y|<y_c$ the space is $\a$ and the  
warp-factor is normalized as $A(y)={\rm exp} (-2Hy)$.
Furthermore, for $|y|>y_c$, the space becomes Minkowskian
and the corresponding warp-factor is a constant, 
$c^2\equiv {\rm exp} (-2Hy_c)$. Thus, at large distances, i.e. $|y|>>y_c$,
the system reduces to a single tensionless brane 
embedded in five-dimensional Minkowski space-time.
For simplicity of presentation in what follows we will deal with 
the positive semi-axis only, i.e., $y \ge 0$ 
(the negative part of the whole $y$ axis is restored
by reflection symmetry). 
Choosing the  traceless covariant  gauge
for graviton fluctuations ($h^{\mu}_{\mu}=0,~~\partial^\mu h_{\mu\nu}=0$)
the  Einstein equations take the form:
\beq
\Psi^{\prime\prime}~-~4H^2\Psi~+~m^2 e^{2Hy}\Psi~=~0,~~~~~~~~0< y < y_c~,
\nonumber \\
\Psi^{\prime\prime}~+~{m^2\over c^2} \Psi ~=~0,~~~~~y>y_c~.
\label{fluct}
\eeq
Where we have introduced the $y$ dependent part of the fluctuations
as follows
$h(x,y) \equiv \Psi(y) {\rm exp}(ipx)$. Furthermore,  the mass-shell
condition for graviton fluctuations  is defined as  $p^2=m^2$.
The equations presented above should be 
accompanied by the Israel matching conditions at the points where the 
branes are located. For the particular case at hand 
these conditions take the form \cite {Rubakov}
\beq
\Psi^{\prime}~+~2H\Psi~=~0, ~~~~~~~y=0~; \nonumber \\
\Psi^{\prime}|_{\rm jump}~=~2H\Psi, ~~~~y=y_c~.
\label{matching}
\eeq 
The solutions to equations (\ref {fluct}) are combinations
of Bessel functions for $0<y<y_c$, and exponentials for
$y>y_c$:
\beq
\Psi_m~(y)~=~A_m ~J_2\left ( {m\over H} e^{Hy} \right )~+~
B_m ~N_2\left ( {m\over H} e^{Hy} \right )~,~~~~~0<y<y_c~, 
\label{Bessel} \\
\Psi_m(y)~=~C_m~{\rm exp} \left ( i{m\over c} (y-y_c) \right)~+~
~D_m~{\rm exp} \left (- i{m\over c} (y-y_c) \right)~,~~~y>y_c~.
\eeq 
The constant coefficients $A_m,~B_m,~C_m$ and $D_m$ 
are determined by using the matching conditions (\ref {matching}) 
(along with the normalization equation). 
The presence of a resonance state 
requires  that the coefficient of the incoming  wave in the solution
($D_m$ in this case)  vanishes at a point in the complex $m$ plane. 
This determines a resonance. 
Calculating $D_m$ and putting $D_m=0$ one finds:
\beq
K_1(\rho)~I_2\left ( \rho e^{Hy} \right )~+~
I_1(\rho)~K_2\left ( \rho e^{Hy} \right )~=~
I_1(\rho)~K_1\left ( \rho e^{Hy} \right )~-~
K_1(\rho)~I_1\left ( \rho e^{Hy} \right )~,
\eeq
where we have introduced a new variable $\rho \equiv -i m/H$. 
This relation can be solved for small values of $\rho$. The result is
$\rho~\simeq~-2 {\rm exp}\left (-3Hy_c\right )~$. 
Therefore, the resonance width is
proportional to
\beq
\Gamma~\propto~ H ~{\rm exp}\left (-3Hy_c \right )~.
\eeq
In the limit $y_c\rightarrow \infty$, the resonance width 
goes to zero and one recovers a zero-mode graviton localized on a positive
tension brane \cite {RS}.

\end{document}